\documentclass[aps,prl,twocolumn,showpacs,letterpaper]{revtex4-1}
\usepackage{graphicx,dcolumn,longtable,epsfig}
\usepackage[usenames]{color}
\usepackage{amssymb}
\usepackage{amsmath}
\usepackage{epstopdf}
\usepackage{bm}

\input epsf.tex
\newcommand{\etal}{{\it et al.}}

\def\jpb#1#2#3{J.~Phys.~B:~{\bf #1},\ #2\ (#3)}

\def\jcp#1#2#3{J.~Chem.~Phys.~{\bf #1},\ #2\ (#3)}
\def\cpl#1#2#3{Chem.~Phys.~Lett.~{\bf #1},\ #2\ (#3)}

\def\pra#1#2#3{Phys.~Rev.~A~{\bf #1},\ #2\ (#3)}
\def\prb#1#2#3{Phys.~Rev.~B~{\bf #1},\ #2\ (#3)}

\def\rmp#1#2#3{Rev.~Mod.~Phys.~{\bf #1},\ #2\ (#3)}
\def\prl#1#2#3{Phys.~Rev.~Lett.~{\bf #1},\ #2\ (#3)}

\def\nat#1#2#3{Nature~{\bf #1},\ #2\ (#3)}

\def\njp#1#2#3{N.~J.~Phys.~{\bf #1},\ #2\ (#3)}
\def\natp#1#2#3{Nat.~Phys.~{\bf #1},\ #2\ (#3)}
\def\lpl#1#2#3{Las.~Phys.~Lett.~{\bf #1},\ #2\ (#3)}
\def\apl#1#2#3{App.~Phys.~Lett.~{\bf #1},\ #2\ (#3)}
\def\jphys#1#2#3{J.~Phys.:~Condens.~Matter~{\bf #1},\ #2\ (#3)}
\def\cjc#1#2#3{Can.~J.~Chem.~{\bf #1},\ #2\ (#3)}
\def\etal{{\it et al.}}
\def\bea{\begin{eqnarray}}
\def\eea{\end{eqnarray}}
\def\be{\begin{equation}}
\def\ee{\end{equation}}

\usepackage{color}

\begin{document}
\title{A microscopic model of electronic field noise heating in ion traps}

\author{A. Safavi-Naini$^{1,2}$, P. Rabl$^3$, P. F. Weck$^4$, and H. R. Sadeghpour$^2$}
\affiliation{$^1$ Department of Physics, Massachusetts Institute of Technology, Cambridge, MA 02139}
\affiliation{$^2$ ITAMP, Harvard-Smithsonian Center for Astrophysics, Cambridge, Massachusetts 02138}
\affiliation{$^3$ Institute for Quantum Optics and Quantum Information of the  Austrian Academy of Sciences, 6020 Innsbruck, Austria}
\affiliation{$^4$ Department of Chemistry \& Harry Reid Center, University of Nevada,
Las Vegas, NV 89154}

\date{\today}

\begin{abstract}
Motional heating of ions in micro-fabricated traps is one of the open challenges hindering experimental realizations of large-scale quantum processing devices. Recently a series of measurements of the heating rates in surface-electrode ion traps characterized their frequency, distance, and temperature dependencies, but our understanding of the microscopic origin of this noise remains incomplete. In this work we develop a theoretical model for the electric field noise which is associated with a random distribution of adsorbed atoms on the trap electrode surface. By using  first principle calculations of the fluctuating dipole moments of the adsorbed atoms we evaluate the distance, frequency and temperature dependence of the resulting electric field fluctuation spectrum. Our theory reproduces correctly the $d^{-4}$ dependence with distance of the ion from the electrode surface and calculates the noise spectrum beyond the standard scenario of two-level fluctuators by incorporating all the relevant vibrational states. Our model predicts a regime of $1/f$ noise which commences at roughly the frequency of the fundamental phonon transition rate and a thermally activated noise spectrum which for higher temperatures exhibits a crossover as a function of frequency.
\end{abstract}


\maketitle

\section{Introduction}
Laser cooled trapped ions represent  one of the most promising systems for the implementation of large scale quantum information processing \cite{HaeffnerPhysRep2008,WinelandLaserPhysics2011}. Most of the basic requirements for building a quantum computer -- the so-called  DiVicenzo criteria~\cite{DiVincenzo2000} --  have been demonstrated in the lab and the generation of entangled states of up to 14 ions has been achieved~\cite{MonzPRL2011}. Many experimental efforts are now focused on the development of miniaturization and micro-fabrication techniques for ion traps~\cite{SeidelinPRL2006,PearsonPRA2006,Chuang2008,AminiNJP2010,AllockNJP2010,DaniilidisNJP2011,Allock2011}, to realize more efficient and also fully scalable quantum computing architectures~\cite{KielpinskiNature2002,AminiNJP2010}.  However, when devices are miniaturized, physics at the short-distance becomes a challenge. This is evident in measurements of Casimir force  (of either sign) \cite{Lamoreaux,Capasso}, or of non-contact friction \cite
{Kuehn,StipePRL2001, Volokitin}, and in the case of trapped ions manifests itself in the appearance of an excess (``anomalous") heating rate as the trap-surface distance is decreased~\cite{Turchette,IonTrapCooling,IonHeating3,IonHeating2,Chuang2008,MITsupercond11,DaniilidisNJP2011,Allock2011}. Therefore, a  detailed understanding of the origin of this noise will be essential for the future progress of trapped ion quantum computing, as well as  the development of several hybrid quantum computing approaches where e.g. ions~\cite{TianPRL2004,DaniilidisJoPB2009}, Rydberg atoms~\cite{SorensenPRL2004}, polar molecules~\cite{AndreNatPhys2006} or charged nano-mechanical resonators~\cite{RablNatPhys2010} are operated in the vicinity of solid state systems.

Theoretical attempts to explain the noise-induced heating of trapped ions -- its distance, frequency and temperature dependence -- have been largely phenomenological. The most common noise source in conductors, the Johnson-Nyquist noise from the trap electrodes or circuitry, has a frequency independent spectrum and decays as $d^{-2}$ with increasing trap-surface distance $d$.  Experimental data, however, is consistent with a $d^{-4}$ scaling for a large variety of trapping geometries (see e.g. Ref.~\cite{DaniilidisNJP2011} for a recent review) and suggests a $1/f$ or even stronger variation over the observed range of frequencies \cite{IonHeating2,Chuang2008}. Therefore, since the early work of Turchette \etal~\cite{Turchette}, the influence of randomly oriented dipole domains (patch potentials) %as the precursor of noisy electric field at the position of the ions, 
has been recognized as the leading source for motional heating of ions. The main assumption of the patch potential model, namely that the electric noise originates from uncorrelated  sources, explains correctly the observed distance dependence of the heating rate, but the model does otherwise not provide further insight  into the physical origin of these fluctuations. More recent experiments with superconducting ion traps~\cite{MITsupercond11} strongly support the fact  that the source of anomalous heating is not in the bulk, but is a surface phenomenon,  and detailed temperature studies \cite{IonHeating3,Chuang2008} reveal that thermally activated processes are at play.

In this work, we develop a microscopic theory of anomalous heating in ion traps and other charged systems. The central assumption in this work is that the electric field noise in ion traps is produced by randomly-distributed fluctuating dipoles on the gold electrode surface, which in turn are formed by surface adsorption of atomic impurities, from the atmosphere or in fabrication. This assumption is in agreement with the experimental observations mentioned above and is supported by the fact that in many traps the heating rate increases over time, especially in the trap loading zone~\cite{DaniilidisNJP2011}. 
 
 In our model, fluctuations of the adatom dipole moment arise from phonon induced transitions between multiple bound surface states. We use analytic models supported by exact density functional calculations to analyze the adatom surface interaction potentials and the resulting time variations of the induced dipole moment.  From this analysis we obtain the electric field fluctuation spectrum, and thereby the ion heating rate, as a function of the relevant microscopic parameters of the atom-surface interaction. Our calculations go beyond the standard scenario of two level fluctuators~\cite{Dutta,martinis,Chuang2008,DaniilidisNJP2011} and we show that the inherent multi-level structure of the surface potential leads to a characteristic frequency and temperature dependence, where even for a single atom a region with $1/f$ scaling emerges due to a distribution of different transition rates. The predicted distance dependence and heating rates are in good agreement  with experimental measurements. More importantly,  the characteristic features of our noise model could  provide more insight into the  microscopic origin of anomalous heating and be tested against other potential mechanisms~\cite{Diffusion}.

 \section{Anomalous heating of trapped ions}

\begin{figure}
\begin{center}
\includegraphics [width=0.5\textwidth]{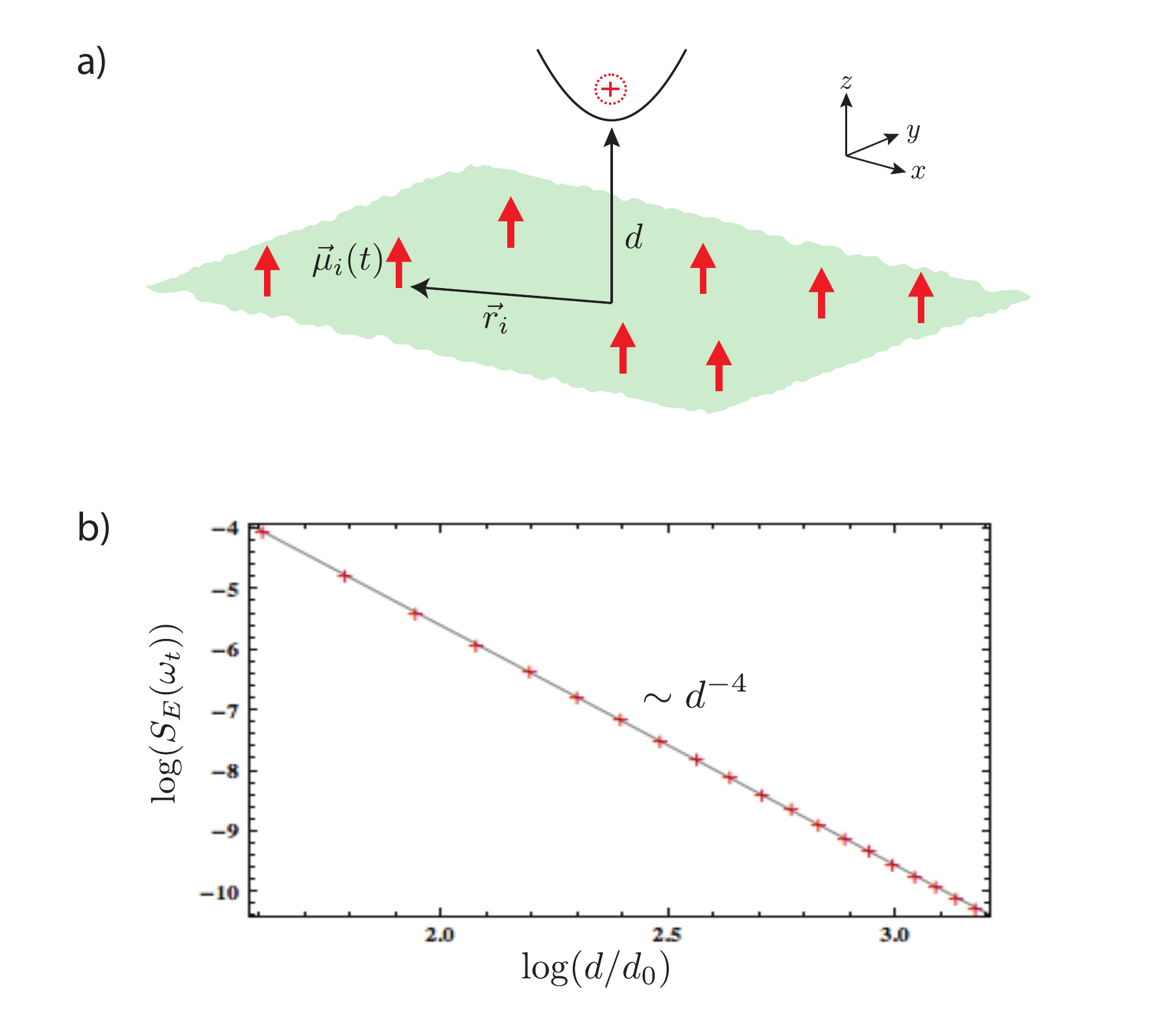}
\caption{ (Color online) a) Fluctuating dipoles associated with a random distribution of adsorbed atoms lead to heating of an ion trapped at distance $d$ above the surface. b) Dependence of the electric field fluctuation spectrum $S_E(\omega_t)$ (in arbitrary units) as a function of the trap-surface distance $d$. The results have been obtained from a numerical simulation of the electric field from $N=100$ uncorrelated dipoles distributed randomly over an area of $100\times 100 \times d_0^2$. $d_0$ is the minimum separation between two dipoles.  
%{\bf In Fig. 5 of Ref. \cite{IonHeating4}, a compilation of several measurements of the distance dependence of the ion noise spectrum are shown and generally fall on the left of our $d^{-4}$ curve. The comparison in magnitude with some of the measurements is generally good.}}
%\PR{check numbers for simulation and figure labels}
}
\label{fig:DipolesOnSurface}
\end{center}
\end{figure}

Fig.~\ref{fig:DipolesOnSurface} (a) shows a typical experimental setting where a single ion of mass $m_I$ and charge $q$ is trapped at a distance $d$ above a metal surface  using combined static and inhomogeneous rf electric fields (Paul traps).  For various designs of surface patterned micro traps~\cite{SeidelinPRL2006,PearsonPRA2006,Chuang2008,AminiNJP2010,AllockNJP2010,DaniilidisNJP2011,Allock2011},
% \textbf{ ANY MORE SUGGESTIONS HERE? ARGHAVAN} 
the separation $d$ is between $20$ to $100\mu$m.    
When the ion is laser cooled the  vibrational ground state fluctuating electric fields couple to the motion of the ion and lead to an increase of the average vibrational occupation number $\bar n$ with a characteristic rate \cite{Turchette} 
%\textbf{ I had to change the notation for below. ARGHAVAN}
 \begin{equation}\label{eq:HeatingRate}
\dot{\bar{n}}=   \frac{q^2}{2m_I \hbar \omega_t} S_E(\omega_t).
\end{equation}  
Here $\omega_t$ is the trapping frequency of the ion, typically in the range  $\omega_t/2\pi\approx  0.1-10$ MHz, and $S_E(\omega)=\int_{-\infty}^\infty d\tau \langle \delta  E(\tau) \delta E(0)\rangle  e^{i\omega \tau}$ is the spectrum of the fluctuating electric field $\delta E(t)$ (projected onto the trapping axis) at the position of the trap. The heating rate~\eqref{eq:HeatingRate} can be measured in experiments and thereby provides an accurate probe of the electric field noise over the accessible distance and frequency range.

%In the vicinity of the surface various different sources of noise are at play. A common noise source is the thermal Johnson-Nyquist noise \cite{Dutta} associated with a finite resistance $R$ of the trap electronics.  The voltage fluctuations $S_V(\omega)$  follow from the fluctuation dissipation theorem and lead to an electric field noise $S_E(\omega) \propto S_V(\omega) d^{-2}\sim k_BT R(\omega) d^{-2}$ at the position of the ion.  The spectrum of Johnson-Nyquist noise is frequency independent (white noise) and since Ohmic systems behave $R (T) \propto [1+\alpha(T-T_0)]$, the temperature dependence scales $S_E \sim T^2$.  
%
%%However, these predictions from thermal voltage fluctuations are inconsistent with experimental observations of the distance, temperature and frequency dependence of the heating rate $\Gamma_h$ \cite{Turchette,Chuang2008}, and other, anomalous heating mechanisms must be considered.
%
%The concept of fluctuating microscopic "patch potential" noise was first advanced by Turchette \etal \cite{Turchette} to derive the distance dependence of the electric field noise. The idea is that each patch is a disk with a radius much less than the radius of the spherical electrode conducting shell, or in the case of ion traps, the distance from the electrode to the ion. These patch potentials are a collection of randomly distributed fluctuating dipoles, due to either impurities adsorption on the surface from the atmosphere or implanted during fabrication.

In our model we consider electric fields originating from a distribution of fluctuating dipoles $\vec \mu_i(t)$ which are associated with individual atoms adsorbed on the surface at positions $\vec r_i$. In accordance with previous ``patch potential" models~\cite{Turchette,CorrelationLength}  the assumption of uncorrelated noise sources leads to  the expected distance dependence $d^{-4}$. This is illustrated in Fig.~\ref{fig:DipolesOnSurface} (b) where we have numerically evaluated the electric field noise of $N=100$ randomly distributed dipoles on a surface. More explicitly, by averaging over a homogenous distribution of atoms with area density $\sigma$ the electric field noise spectrum at the position of the ion can be written as
\begin{equation}\label{eq:SE}
S_E(\omega_t)= \frac{3}{8} \frac{\sigma}{(4\pi \epsilon_0)^2} \frac{ S_\mu(\omega_t)}{d^4},
\end{equation}     
where $S_\mu(\omega)=\int_{-\infty}^\infty d\tau \langle \delta \mu_z (\tau) \delta \mu_z(0)\rangle  e^{i\omega \tau} $
% {\bf we have two definitions for $S_E(\omega)$, see also Eq. 1. Do we need both?. I think the other relation is more useful for people making measurements. Peter feel free to cut out something here. ARGHAVAN}
 is the spectrum of the fluctuating dipole of a single adatom. %Using  Eq.~\eqref{eq:HeatingRate} and Eq.~\eqref{eq:SE} as a starting point 
Our main goal in the remainder of this paper is to provide a microscopic derivation of the dipole-fluctuation spectrum $S_\mu(\omega)$, which by using Eq.~\eqref{eq:HeatingRate} and Eq.~\eqref{eq:SE} allows us to establish a direct relation between the ion heating rate and the microscopic details of the atom surface interactions.

\section{Atom-surface interaction}

Atoms  approaching a surface experience an attractive force which at large distances is the well-studied van der Waals (vdW) potential $\sim - C_3/z^3$ which eventually becomes repulsive again when the electronic wavefunctions of the adsorbant and the bulk atoms overlap. An atom approaching the surface can lose energy by phonon induced processes and get trapped in the resulting potential well.  We develop a model, based on physical intuition, which captures the essential aspects of the atom-surface interaction.  To this end, we present {\it ab initio} density-functional calculations (DFT) of hydrogen adsorption on Au(111) surface and obtain the interaction potential normal to the surface. This potential is used as proxy for our model atom-surface potentials which more closely mimic realistic impurity reactivity on gold surfaces. The utility in using a parametric model potential rests in its flexibility for tuning phonon transition frequencies, and range of both short-range and vdW interactions. 

%Another clear advantage in using surface potentials is to investigate multi-level schemes.

\subsection{Ab initio atom-surface potentials}

For the calculations to be manageable, we chose hydrogen adsorption on the gold surface. Clearly, surface contaminants are more than hydrogen atoms and contain atmospheric species. We incorporate such species by devising realistic interaction potentials \ref{model} to model the adsorption of atmospheric or fabrication impurities on gold surfaces. For the H--Au interaction, all-electron scalar relativistic calculations of the total energy and
optimized geometries of a H--Au surface model system were performed using the spin-polarized density functional theory as implemented in the DMol3
software \cite{delley2000}. The exchange correlation energy was calculated using the local gradient approximation (LDA) with the parametrization 
of Perdew and Wang (PWC) \cite{perdew1992}. 

\begin{figure}
%\begin{center}
%\includegraphics[width=0.4\textwidth]{HAu.pdf} 
\includegraphics[width=0.4\textwidth]{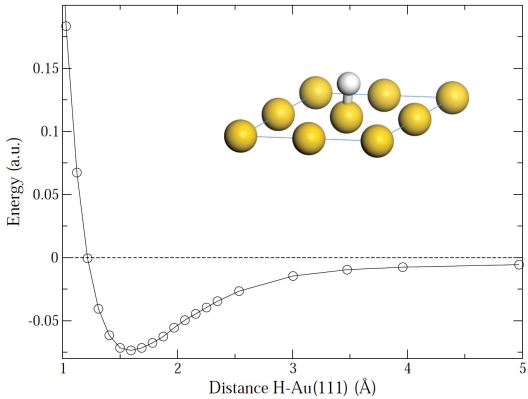} 
%\end{center}
\caption{\label{HAu} 
(Color online) Potential energy curve for a H atom interacting with a $2\times2$ Au(111) 
monolayer along the normal to the surface calculated using density functional 
theory within the local density approximation.}
\end{figure}

In Fig. \ref{HAu}, the H-Au(111) surface potential is shown.  The long-range interaction of atom-conductor potential is given by $C_3/z^3$, where $C_3$ is proportional to the atomic electric dipole transitions, and $z$ is the normal to the surface. The coefficient $C_3=\frac{1}{4\pi}\int_0^\infty \alpha(i\omega)d\omega $ can be obtained numerically by evaluating the dynamic atomic polarizability at imaginary frequencies. The value of $C_3$ for hydrogen is given in Ref. \cite{babb} as $C_3=7.36\times 10^{-5}$ a.u. Allowance is made for the fact that Au dielectric constant is not infinite, i. e. $C_3(Au)=\frac{\epsilon-1}{\epsilon+1}C_3$, where $\epsilon$ is the dielectric constant for Au. The fundamental transition frequency is more than 25 THz, many orders of magnitude larger than any frequency scale in the experiments. Heavier adatoms on the gold surface mass scale the interaction potentials, leading to lowering of the transition frequencies. 
%Double numerical basis sets including polarization functions on all atoms (DNP) were used in the calculations. The DNP basis set corresponds to a double-$\zeta$ quality basis set with a $p-$type polarization function added to hydrogen and $d-$type polarization functions added to heavier atoms. The DNP basis set is comparable to 6-31G$^{**}$ Gaussian basis sets \cite{hehre1986} with a better accuracy for a similar basis set size \cite{delley2000}. In the generation of the numerical basis sets, a global orbital cutoff of $4.5$ \AA~was used.
%The energy tolerance in the self-consistent field calculations was set to $10^{-6}$ Hartree. Optimized geometries were obtained using the direct inversion in a subspace method (DIIS) without symmetry constraints with an energy convergence tolerance of $10^{-5}$ Hartree and a gradient convergence of $2\times 10^{-3}$ Hartree/Bohr.

\subsection{Model atom-surface potentials}
\label{model}
\begin{figure}
\includegraphics [width=0.5\textwidth]{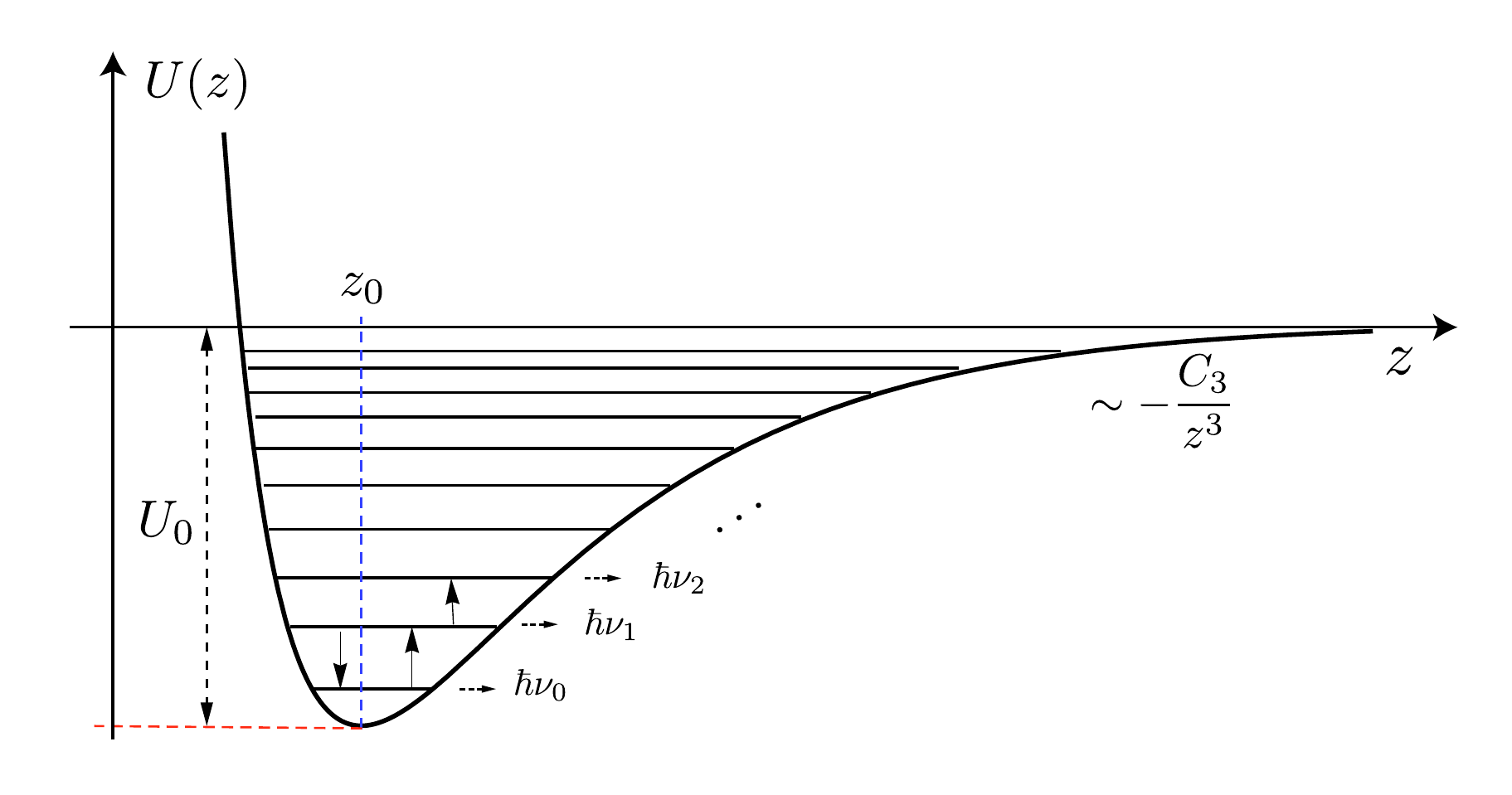}
\caption{(Color online) Surface potential with potential depth $U_0$ and minimum at $z_0$. The potential support several bound vibrational states with energies $\hbar \nu_i$.  The arrows represent the phonon mediated transitions between the bound states. The parameters used for this plot are $U_0=12$ meV, $z_0=6.05a_0$, $\beta=0.95 a_0^{-1}$ and an atomic mass of $m=20$ a.u. These parameters correspond to the case of Ne on a gold surface. } 
\label{potential}
\end{figure} 
To describe the atom-surface interactions for wider range of atoms we use a model potential which is commonly referred to as exp-3 \cite{hoinkes}. This potential is frequently used in surface science studies \cite{hoinkes} and provides a suitable description for short and medium range distances. Our potential in this range is described by
\begin{equation}\label{Uz}
U(z)= \frac{\beta z_0}{\beta z_0  - 3}
  U_0 \left[\frac{3}{\beta z_0} \,e^{\beta z_0 (1-z/z_0)} -\left( \frac{z_0}{z}\right)^3\right],
\end{equation} 
where $z_0$ is the equilibrium position, $U_0$ is the depth of the potential and $\beta$ is the reciprocal range of repulsion.  The typical shape of this potential  is plotted in Fig.~\ref{potential}. In the long range Eq.~\eqref{Uz} reproduces the correct $\sim - C_3/z^3$ form where $C_3=\frac{\beta \,z_0^4}{\beta z_0 - 3}U_0$ in terms of the model parameters, and by adjusting these three parameters, we can modify  $U(z)$ to fit realistic atom-surface potentials. 
%The typical shape of this potential together with the energies of the lowest bound vibrational states is shown in Fig.~\ref{potential}. 
The potential supports several bound vibrational states with energies $E_i=\hbar \nu_i$ as indicated in Fig.~\ref{potential}.
By using an harmonic expansion of $U(z)$ we find that for an adatom of mass $m$, the typical vibrational excitation frequency is approximately given by
\begin{equation}
\nu_{10}:=\nu_1-\nu_0\approx  \sqrt{\frac{U_0}{m z_0^2}\frac{3(\tilde \beta^2-4\tilde \beta)}{( \tilde \beta -3)} },
\end{equation}
where $\tilde \beta=\beta z_0$. From this result, we also find a rough estimate of the number of strongly bound vibrational states $N_{b}\approx U_0/(\hbar \nu_{10})$.

From the exact H-Au potential shown in Fig. \ref{HAu}, we deduce $U_0\approx 2$ eV, corresponding to a temperature of $T=1.6\times10^4$ K, $z_0\approx 1.6 \AA$  and $\beta\approx 3.91 \AA^{-1}$, as example for an adatom with a high reactivity with the Au surface. Because of the low mass of H, also the vibrational frequencies are in the range of $\nu_{10}/2\pi \approx 40$ THz. In general we expect a similar reactivity for other alkali atoms as can be seen for example for K-Ag where $U_0=1.79 eV$ and $z_0 \approx 2 \AA$ \cite{PotassiumData}. However, due to the larger mass we obtain significantly lower vibrational frequencies  $\nu_{10}/2\pi \approx 4$ THz. In contrast, for weakly interacting atoms, we find from the widely-studied noble gas-metal potentials (see for instance, Refs. \cite{dasilva,Ossicini} ) that potentials can be wider and much shallower.  For example, for Ne on a gold surface $U_0\approx 0.012$ eV,   $z_0\approx 3.1 \AA $ and 
$\beta\approx 1.86 \AA^{-1}$.  Within this typical range of potential parameters and adatom mass $m\sim 10-150$ a.u., we expect the relevant  vibrational frequency scales to be in the  $\nu_{10}/2\pi \approx 0.1-1$ THz regime. For our model potential in Fig.~\ref{potential}, and using the reduced mass of Ne-Au, we find $\nu_{10}/2\pi \approx 0.3$ THz.

In our model of atom-surface interactions we ignore the dependence of $U(z)$ on $x$ and $y$ due to surface roughness. However, for a metal surface, where the electrons are smeared out, this variation should be weak.  Also, the random motion of an otherwise fixed dipole along the $xy$plane would lead to a different,  $d^{-6}$ scaling of the heating rate which is not supported by experimental data.

\section{Fluctuating Dipoles}
%Based on the 1D interaction potential $U(z)$ in Fig. \ref{fig1} 
\subsection{Adatom dipoles}

\begin{figure}[htp]
\includegraphics [width=0.5\textwidth]{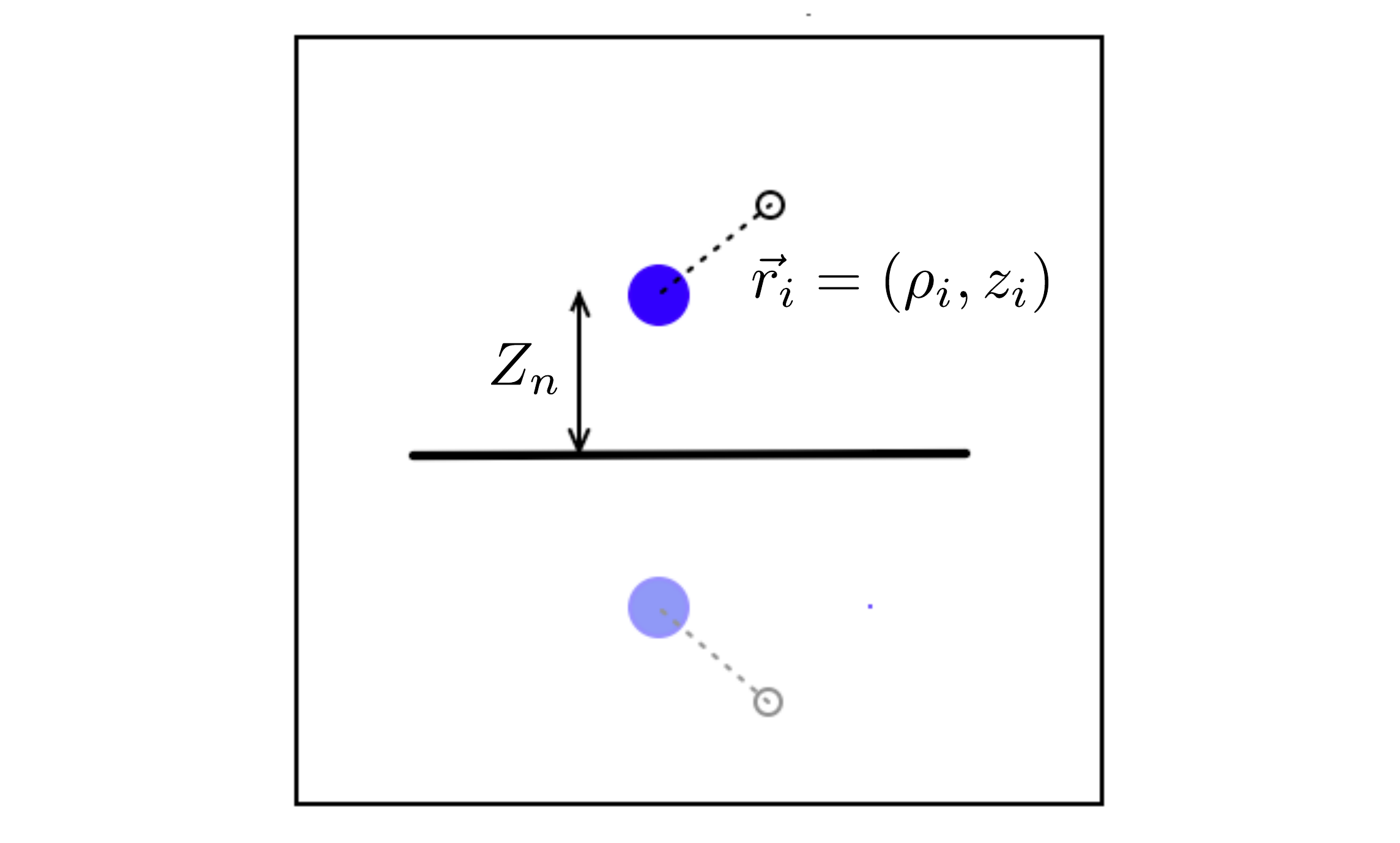}
\caption{(Color online) The adatom nucleus is at a distance $Z_n$ from the surface. The electron is at a distance $\vec{r}_i$ from the nucleus. }\label{image}

\end{figure}

Adatoms adsorbed on the surface exhibit a finite permanent dipole moment perpendicular to the surface. This induced dipole moment  can be understood from the  distortion of the electronic wavefunctions.  It is commonly argued that the dipoles form, when the impurity adatom valence electrons penetrate into the surface conduction bands, modifying the surface work function. Here, we calculate the magnitude of the electric dipole moments of the adatoms, by resorting to elementary electrostatics, using image charge techniques. The electrons and the ionic core interact with the surface electrons through their respective image charges, as depicted in Fig. \ref{image}. We begin by writing the potential seen by the orbital electron due to its image charge \cite{Anton},
%\begin{eqnarray}
\be
V=-\frac{e^2}{8Z_n^3}\sum_{i} (z_i^2+ \frac{1}{2}\rho_i^2)  
-\frac{3e^2}{16Z_n^4}\sum_{i} (z_i^3+ \frac{1}{2}z_i \rho_i^2),
\label{Vz}
\ee
%\end{eqnarray},
where $\vec{r}_i=(\vec{\rho}_i, z_i)$ is the position of the $i^{th}$ electron and $Z_n$ is the distance from the nucleus of the adatom to the metallic surface (see Fig. \ref{image}). In specifying the position of the electron, we use $z_i$ to refer to the distance from the electron to the nucleus along the normal to the surface. The first term is the above-discussed vdW interaction with the surface. The second term in Eq. \eqref{Vz} vanishes in the first order of perturbation expansion, but contributes to the second order energy shift, $\Delta E ^{(2)} $.

\begin{figure}
\includegraphics [trim =10mm 10mm 0mm 25mm, clip,width=0.45\textwidth]{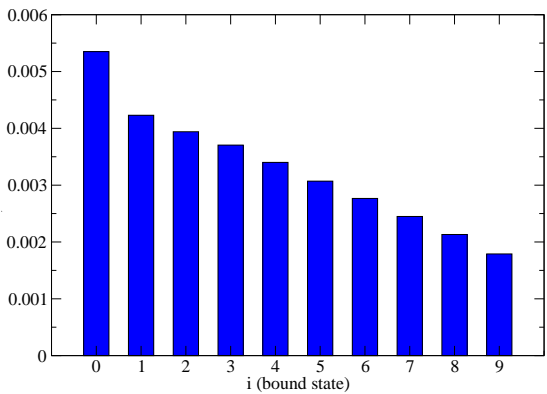}
\caption{(Color online) Magnitude of the average induced dipole moments $\mu_{z,i}$ for different vibrational states $|i\rangle$. The values are shown for the model potential parameters given in Fig.~\ref{potential} and a polarizability of $\alpha({\rm Ne})\approx 0.36\AA^3$. }\label{mu}
\end{figure}

Following Ref. \cite{Anton}, we use a variational ansatz of the form $\phi =(1+\sum_i \lambda_i z_i)\psi_0$, where $\psi_0$ are the unperturbed atomic wave functions, and calculate $\Delta E^{(2)} $, by 
%\begin{equation*}
%\phi =(1+\sum_i \lambda_i r_i)\psi_0
%\end{equation*}
varying the parameters $\lambda_i$. From the deformed electronic wavefunction $\phi$, we find an approximated induced dipole moment of the atom near the surface, $P(Z_n)=\langle \phi \vert \sum_i e \vec{r}_i \vert \phi \rangle$. For hydrogen atom $P(Z_n)= \frac{4.5 ea_0^5}{Z_n^4} $, where $a_0$ is the Bohr radius. Since the numerical factor in the expression is just the static polarizability of hydrogen, $\alpha_H=4.5\,a_0^3$, this expression can be generalized to a generic atom with polarizability $\alpha$,
\begin{equation}\label{eq:dipole}
P(Z_n)=0.47ea_0^{1/2}\alpha^{3/2}\frac{1}{Z_n^4}.  
\end{equation}
%To find the induced moment corresponding to each bound state of an adatom surface potential, we use the model potential shown in figure \ref{potential}. 
We use the numerically constructed wave functions $\vert i \rangle$, corresponding to the vibrational bound states of our model potential $U(z)$, to evaluate the  average induced dipole moments 
\begin{equation}
\mu_{z,i}:=\langle i | P(z) | i \rangle.
\end{equation}
%To check the accuracy of this calculation we evaluate 
We should note that due to the image charges the dipole moment seen by the atom will be twice the induced dipole moment in Eq.~\eqref{eq:dipole}. In Fig.~\ref{mu} we plot the resulting induced dipole moments $\mu_{z,i}$ for the model potential parameters shown in Fig.~\ref{potential}.  These parameters represent a weakly bound adatom similar to the Ne-Au surface potential \cite{ Ossicini}. Using this potential together with the polarizability of Ne, $\alpha({\rm Ne})=0.36 \AA^3$ \cite{Kumar}, and $Z_n=z_0$, we obtain an induced ground state dipole moment of $\mu_{z,0}({\rm Ne})=0.005$D. Hence the dipole moment seen by the Ne atom is approximately $0.01$ D which is in good agreement with $0.016$ D calculated for Ne in \cite{dasilva}. Note that $\mu_{z,i}\sim \alpha^{3/2}/z_0^4$ and in general the typical magnitude of induced dipole moments is $\sim 1$D. For example in \cite{Volokitin, Senet1999}, it is shown that the dipole moment for Cs absorbed on Cu(100) is 4D. Similarly in \cite{Russier} the induced dipole moments for K adsorbed on W, Ni and Pd ranges from 1.45 to 3.1 D. 

\subsection{Phonon induced transition rates}

The emission and absorption of phonons in the bulk lead to transitions between different bound vibrational states  and result in fluctuations of the induced dipole moment of the adatom.  
To evaluate the phonon induced transitions rates, 
% in the potential in Fig. \ref{potential} can be calculated using Fermi's golden rule. The model potential in Fig. \ref{fig 2} holds 10 vibrational levels {\bf the parameters for the potential should be given in the figure caption}.  Here, we describe how we calculate the phonon transition rates. These rates determine the time evolution of states, when incorporated in a density matrix formalism of level dissipation.
we approximate the trap electrode surface by a semi-infinite crystalline slab. The position of an atom in the solid is given by
$$\mathbf{r}_i=\mathbf{r}_i^0+\mathbf{u}_i,$$
where ${\bf r}_i^0$ is the equilibrium position of the $i^{th}$ atom and ${\bf u}_i$ is its deviation from the equilibrium. We can write the displacements $\mathbf{u}_i$ terms of bosonic operators $a_\lambda(\mathbf{q})$ for the phonon eigenmodes, 
\be \label{eq:PhononModes}
{\bf u}_i=\sum_{\mathbf{q},\lambda} \sqrt{\frac{\hbar}{2NM\omega_{\lambda,{\bf q}} }}\vec{\epsilon_\lambda}(\mathbf{q}) (a_\lambda(\mathbf{q})e^{i\mathbf{qr_i^0}}+h.c.),
\ee
where ${\bf q}$ is the quasi-momentum, $N$ is the number of atoms in the bulk and $M$ is their mass.  For each ${\bf q}$ the normalized vectors $\vec{\epsilon_\lambda}(\mathbf{q})$ describe the three orthogonal phonon polarizations.

In the presence of phonons, the adatom surface potential $U(z)$ which in Eq. (3) has been defined for a static surface, will in general depend on the fluctuating positions of the bulk atoms so that $U(z)\rightarrow U(z, \{\mathbf{r}_i\} )$.  
%To account for small position fluctuations of the bulk atoms we must  
%We want to find the single-phonon transition rates between two adjacent bound states in the model potential shown in Fig. \ref{potential}. The bound state energies are $\vert E_i\rangle$ and $\vert E_f \rangle$. 
Expanding this potential to first order in ${\bf u}_i$ gives,
\begin{equation}\label{eq:Uzr}
U(z, \{\mathbf{r}_i\} )\simeq U(z,\{\mathbf{r_i}^0\} )+ \sum_i \nabla U(z,\{\mathbf{r_i}^0\} ){\bf u}_i.
\end{equation}
Since our potential is already averaged over the two transverse directions, we are only interested in the variations in the normal direction. Further, the dominant deformation of the potential arises from the closest surface atom and in the following, we restrict the sum in Eq.~\eqref{eq:Uzr} to a single atom. 
%Assuming $\Delta E= E_i-E_f > 0$, where $E_{i,f}$ are the vibrational energies, 
We use Fermi's Golden Rule to evaluate the phonon induced transition rate between two vibrational states $|i\rangle$ and $|f\rangle$,
%\begin{equation}
%\begin{split}
%\Gamma_{i\to f}=\frac{2\pi}{\hbar} \sum_{\mathbf{q},\lambda} &\big| \langle n_{{\bf q}}+1|  \langle f\vert \frac{dU}{dz}\cdot u_z\vert i \rangle|n_{{\bf q}}\rangle \big|^2  \delta(\Delta E -\hbar \omega_{\bf q})\\
%&\big| \langle n_{{\bf q}}-1|  \langle f\vert \frac{dU}{dz}\cdot u_z\vert i \rangle|n_{{\bf q}}\rangle \big|^2  \delta(\Delta E +\hbar \omega_{\bf q}),
%\end{split}
%\end{equation} 
\begin{equation}\label{eq:GoldenRule} 
\begin{split}
&\Gamma_{i\to f}=\frac{2\pi}{\hbar} \sum_{\mathbf{q}} \big| \langle f\vert \frac{dU}{dz}\vert i \rangle\big|^2  
\Big( \big| \langle n_{{\bf q}}+1|   u_z |n_{{\bf q}}\rangle \big|^2 \times \\ &\delta(\Delta E -\hbar \omega_{\bf q})   +\big| \langle n_{{\bf q}}-1|   u_z |n_{{\bf q}}\rangle \big|^2 \delta(\Delta E +\hbar \omega_{\bf q})\Big), \end{split}
\end{equation} 
where $\Delta E= E_i-E_f $ is the difference between the vibrational energies $E_{i,f}$, 
 $n_{\bf q}$ are phonon mode occupation numbers and in this equation ${\bf q}\equiv ({\bf q},\lambda)$ includes the polarization label.  
Assuming $\Delta E>0$ and using the mode decomposition in Eq.~\eqref{eq:PhononModes}, the resulting phonon emission and absorption rates can be written as 
%and absorption rates are given by Ê
%\begin{equation}
%\Gamma_{i\to f}=\frac{2\pi}{\hbar} g(E) n(E) \vert \langle f\vert \frac{dU}{dz}\cdot u\vert i \rangle \vert^2
%\end{equation}
%where $g(E)=\frac{\Delta E^2}{2\pi^2\hbar^3 v^3}$ is the phonon density of states for a material with speed of sound $v$, $n(E)$ is the Bose factor and $u$ is the displacement amplitude of the lattice atoms. Equating the time average classical energy per unit volume to the energy of the phonon we get 
%$$
%\hbar \omega=2\rho u^2 \omega^2
%$$
%where $\rho$ is the bulk density of the material and $\omega$ is the phonon frequency \cite{phillips1987}.  We can now find the emission and absorption rates as follows:
\begin{eqnarray}
\label{eq:rates1}
\Gamma_{i\to f}&=&\frac{\Delta \omega_{if}}{2 \pi\hbar v^3  \rho}  \vert \langle f \vert \frac{d}{dz} U(z)\vert i \rangle \vert ^2 \left(n(\Delta \omega_{if})+1\right),\\
\label{eq:rates2}
\Gamma_{f\to i}&=&\frac{\Delta \omega_{if}}{2 \pi \hbar v^3 \rho}   \vert \langle f \vert \frac{d}{dz} U(z)\vert i \rangle \vert ^2 n(\Delta \omega_{if}),
\end{eqnarray}
where $\Delta \omega_{if}=|E_i-E_f|/\hbar$, $n(\Delta \omega)=(e^{\hbar\Delta \omega/k_BT}-1)^{-1}$, $v$ is the averaged speed of sound in the surface material and $\rho$ is its bulk density. 

In the following, we denote by $\Gamma_0\equiv \Gamma_{1\to0}(T=0)$, the zero temperature decay rate from the first excited to the lowest vibrational state. From a simple harmonic approximation of $U(z)$ around its minimum at $z_0$, we obtain the  scaling
\begin{equation}
\Gamma_0\approx \frac{1}{4\pi} \times \frac{\nu_{10}^4 m }{v^3 \rho}.
\end{equation}
Using $v=3962$ m/s and $\rho=19.3$ g/cm$^3$ for Au, and the model potential parameters given in Fig. (\ref{potential}), with $\nu_{10}/2\pi=0.3$ THz, we find $\Gamma_0 /2\pi \approx 3.31$ MHz. For K-Au where $\nu_{10}/2\pi=4$ THz $\Gamma_0/2\pi$ is approximately $67$ MHz and in general we expect $\Gamma_0/2\pi$ to range from about 1 to a few hundred MHz. Note that the validity of Eq.~\eqref{eq:rates1} and Eq.~\eqref{eq:rates2} is restricted to transition frequencies $\Delta \omega_{if}$ smaller than the Debye frequency $\omega_D$ of the bulk material, which for gold is about $3.6$ THz.

\subsection{Dipole fluctuation spectrum} 
We are interested in the fluctuation spectrum of the induced dipole moment of a single adatom, defined as  
\begin{equation}
S_\mu(\omega)=\int_{-\infty}^\infty d\tau \left( \langle \mu_z (\tau) \mu_z(0)\rangle -\langle \mu_z(0)\rangle^2\right) e^{i\omega \tau}.
\end{equation}
Summarizing the results from the previous sections we can write the dipole moment operator as
$ \mu_z =  \sum_i   \mu_{z,i} \rho_i$, where  $\rho_i= |i\rangle\langle i| $ is the projection operator on the vibrational level $|i\rangle$. Therefore, for given values of $p_i$ the dipole fluctuation spectrum can be related to the set of two-point correlation functions $\langle \rho_i(t) \rho_j(t+\tau) \rangle$ of the vibrational populations. The populations in turn evolve according to the master equation
\begin{equation}
\frac{d}{dt}\langle \rho_i\rangle=\sum_j M_{ij}\langle \rho_j \rangle,
\end{equation}
where the diagonal $M_{ii}= -\sum_{j \neq i} \Gamma_{i \to j}$  and the off-diagonal elements $M_{ij}=\Gamma_{j\to i}$ are determined by  the phonon induced transition rates discussed above.
 We evaluate  the dynamics of the correlations $\langle \rho_i(t) \rho_j(t+\tau) \rangle$ by first introducing the condition $\sum_i \langle \rho_i \rangle =1$ into the master equation. For N bound states we have,
\begin{eqnarray}
\frac{d}{dt}\langle \rho_i\rangle&=&\sum_{j\neq N} M_{ij} \langle \rho_j\rangle +M_{iN}(1-\sum_{k\neq N} \langle \rho_k\rangle )\nonumber \\
 &=&\sum_{j\neq N} (M_{ij}-M_{iN}) \langle \rho_j\rangle +M_{iN}.
\end{eqnarray}
Since we are only interested in level populations, all coherences, $\rho_{ij}=|i\rangle\langle j| $, in the density matrix can be omitted.  Using the quantum regression theorem, we find for $i<N$,
\begin{equation}
\frac{d}{dt} \langle \rho_i(\tau)\rho_k(0)\rangle
          =\sum_{j\neq N} (M_{ij}-M_{iN}) \langle \rho_j(\tau)\rho_k(0) \rangle +M_{iN} \rho_{k}^{(0)}
\end{equation}
where $\rho_k^{(0)}$ is the steady state population in level $k$.  For $i= N$ we obtain
\begin{equation}
\begin{split}
\frac{d}{d\tau}\langle \rho_N(\tau) \rho_k(0)\rangle =& \sum_{i=1}^{N-1}(M_{Ni}-M_{NN})\langle \rho_i(\tau) \rho_k (0)\rangle \\
&+M_{NN}\rho_k^{(0)}.
\end{split}
\end{equation}
All two-point correlation functions can be calculated from the above two equations, and the full noise spectrum is obtained by summing all the two-point correlations.

\section{Results and discussion }
\begin{figure}
\includegraphics  [scale=0.35]{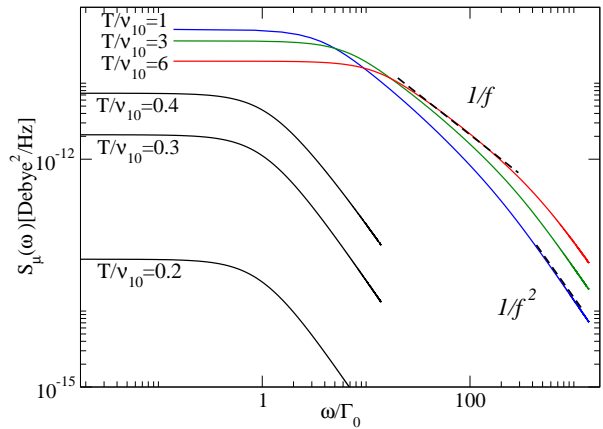}
\caption{ (Color online) The dipole fluctuation spectrum $S_\mu(\omega)$ is plotted for six different temperatures $T$ (given in units of the vibrational excitation frequency $\nu_{10}$) and the model potential parameters given in Fig.~\ref{potential} and Fig~\ref{mu}. The frequency is scaled by the zero temperature transition rate $\Gamma_0$. 
%At $T/\nu_{10}=0.2 and 0.3$ two levels are populated. 
The curves in black ($k_BT/\hbar\nu_{10}=0.2,0.3,0.4$) correspond to temperatures where only the two lowest vibrational states are populated. At these temperatures the spectrum resembles that of a two level fluctuator (see Eq.~\eqref{eq:TLF}) and its size increases with increasing temperature. For higher temperatures (shown in red, green and blue) more and more bound states are thermally occupied and an intermediate regime emerges where the dipole fluctuation spectrum exhibits $1/f$ noise scaling. For all temperatures, the crossover from the flat (white noise) regime to a $1/f^2$ noise or $1/f$ noise scaling occurs at around $\omega_c\approx \Gamma_0(n(\nu_{10})+1)$.  
%
%there is some population in the 3rd level. $T/\nu_{10}=1$ corresponds to 4, $T/\nu_{10}=4$ to 6 levels and and at $T/\nu_{10}=9$ all 10 bound states have significant steady state populations.  When only TLS is involved, the cross over from the white (flat) noise spectrum to a frequency dependent ($1/f^2$) spectrum occurs. For other multi-level excitations (red and green curves), the $1/f^2$ spectrum is preceded by the $1/f$ noise. A crossing of the curves occurs in the $1/f$ spectrum, as the frequency increases. The noise power drops with increasing temperature beyond TLS. For temperatures below the excitation of the TLS  noise power increases with increasing temperature, and the system behaves like a TLS.  
}
\label{fig:spectrum}
\end{figure}

In Fig.~\ref{fig:spectrum}, we plot the typical behavior of the dipole fluctuation spectrum $S_\mu(\omega)$ as a function of frequency. The spectrum has been evaluated for the model potential parameters given in Fig.~\ref{potential} and for different temperatures $T$.   The lowest temperature $k_BT=0.02 U_0$ corresponds to a situation where  the thermal energy is smaller than the vibrational energy $\hbar \nu_{10}\approx 0.1 U_0$  and only transitions between the two lowest vibrational states contribute to dipole fluctuations. In this case, the dipole fluctuation spectrum resembles that of a thermally activated two level system \cite{martinis}, where at low temperatures
 \begin{equation}\label{eq:TLF}
 S_\mu(\omega)\simeq    (\mu_0-\mu_1)^2  \frac{2\Gamma_0}{\omega^2+\Gamma^2_0} \,e^{-\frac{\hbar \nu_{10}}{k_BT}}.
 \end{equation}
For frequencies below the fundamental phonon transition rate, $\omega< \Gamma_0 $, the spectrum is  frequency independent (white noise) while above $\Gamma_0$  the scaling changes to $1/\omega^2$ as one would expect for a two-level system \cite{martinis}.  The noise is thermally activated with a characteristic temperature scale $T=\hbar \nu_{10}/k_B$, which depending on the atomic species ranges from few Kelvin to above room temperature for light and tightly bound adatoms like hydrogen.

When the temperature exceeds the characteristic vibrational energy $T>\hbar \nu_{10}/k_B$, more and more vibrational levels are populated and contribute to the dipole fluctuations (curves in red, green and blue in Fig.~\ref{fig:spectrum}).  An intermediate frequency regime appears where the noise spectrum exhibits a $1/f$ scaling.  We emphasis that in contrast to standard models for $1/f$ noise, based on a random set of two level fluctuators with varying parameters~\cite{Dutta,martinis}, this $1/f$ scaling occurs in our model even for a single dipole and emerges from a distribution of different vibrational transitions rates $\Gamma_{if}$ which contribute to the dynamics.  We find that the approximate crossover between the white noise and the $1/f$ noise regime occurs at $\omega\approx \Gamma_0\times ( n(\nu_{10})+1)$.
To check the validity of our model, we compare our spectrum to the measured values. Using Eq. \ref{eq:SE} we can relate $S_\mu(\omega)$ to $S_E(\omega)$. Since $S_\mu \propto \mu^2$, we note that in the $1/f$ noise region, $10^{-11}<S_\mu (\omega)<10^{-7}$ D$^2$/Hz. The value for $\Gamma_0 \sim 1-10$ MHz, and the $1/f$ behavior sets in at $10-100 \,\Gamma_0$ which corresponds to $10-100$ MHz. Using the coverage fraction $\theta=0.1$ corresponding to about $\sigma \sim10^{18}$ m$^{-2}$, we find $ 3.2\times10^{-8}<\omega S_E(\omega)<0.0032$ V$^2$/m$^2$ at $d_0=10\,\mu$m. The experimentally measured values range between $10^{-7}- 0.001$ V$^2$/m$^2$ \cite{DaniilidisNJP2011}.

%
%
%As more levels reach steady state population (curves in red, green and blue in Fig. \ref{fig:spectrum}), there is a crossing, at which the noise power increases with increasing temperature. This occurs when the $1/f$ behavior sets in and $\omega/\Gamma_0 >> 100$. The inset in Fig. \ref{spectrum}(a) shows a magnified region of this crossing. When more than two levels are populated in steady state, i. e. curves for $T/\nu_{10}=4$ (six levels) (green) and $T/\nu_{10}=9$ (10 levels) (red), $1/f$ noise emerges after the white noise, for $\omega/\Gamma_0 \sim 100$ (10 MHz), and only at much higher frequencies, $\omega/\Gamma_0 \sim 10^4$ (1 GHz), the $1/f^2$ sets in. 

\subsection{Temperature dependence}
\begin{figure}[htp]
\begin{center}
\includegraphics  [width=0.55\textwidth]{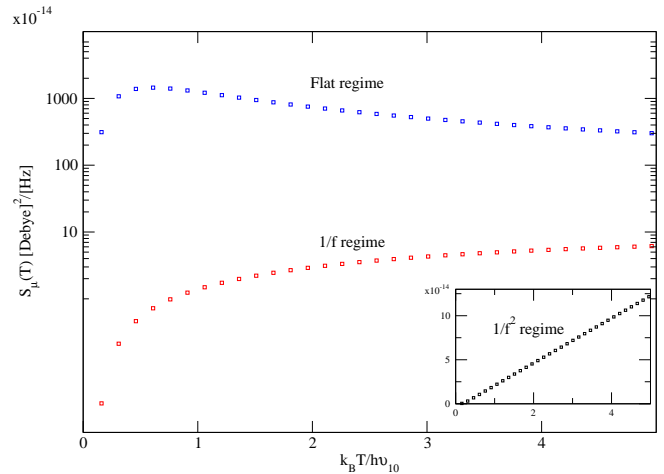} 
\end{center}
\caption{(Color online) Temperature dependence of the dipole fluctuation spectrum vs. normalized temperature ($k_BT/\hbar\nu_{10}$). The three curves show the temperature dependence evaluated for frequencies in the  ``white noise regime" ($\omega\rightarrow 0$), the  ``$1/f$ regime" ($\omega=20\times \Gamma_0$) and the high frequency, ``$1/f^2$ regime" ($\omega=100\times\Gamma_0$). The other parameters are the same as in Fig.~\ref{fig:spectrum}. }
\label{temp}
\end{figure}

 The temperature dependence of the ion heating noise is more succinctly displayed in Fig. \ref{temp} as a function of scaled temperature in units of $\nu_{10}$. At low frequencies, i.e. in the white noise regime, the fluctuations are thermally activated, showing a peak at $k_B T/\hbar \nu_{10}\approx 1$, while for higher temperature they fall off again as $\sim 1/T$.
The suppression of low frequency fluctuations with increasing temperature is in principle expected from a single two level fluctuator. In some models based on multiple two level systems with a distribution of activation energies noise increases with increasing temperature over the whole frequency range \cite{martinis}. In our model this  behavior is recovered for frequencies within the ``1/f region" of the spectrum, $\omega\approx 20 \times \Gamma_0$ , where the temperature dependence matches some of the experimental findings described in \cite{Chuang2008} and fits an Arrhenius curve $F(T)= S_Te^{-T_0/T}$  with the parameters, $S_T=10^{-13}$ D$^2$/Hz and $T_0\approx0.242\,U_0=56.9$ K. Finally, for very large frequencies, $\omega \gg \Gamma_0$, i.e. in the $1/f^2$ regime, noise scales linearly with T.  

%To characterize the temperature dependence of the spectrum we look at three different regions: flate (white noise), $1/f$ and $1/f^2$. For each region we picked a frequency and looked at the magnitude of noise as a function of temperature. The results are shown in Fig. \ref{temp}. In the $1/f$ region we considered the two models described in \cite{jarek}. The Arrhenius curve $F(T)= S_Te^{-T_0/T}$  with $S_T=2.45\times10^{-12}$ and $T_0= 3.83U_0$ was the best fit to our data. The data and the fit curve are shown in Fig. \ref{temp}(b).

\section{Summary and conclusions}
In summary, we have developed a microscopic model for electric field noise generated by fluctuating dipoles  associated with adatoms on a metallic surface.  We have shown that phonon induced transitions between different bound vibrational states cause fluctuations of the induced dipole moments and generate electric field noise which  can contribute to the anomalous heating observed in surface ion traps. The analysis presented in this work has been largely based on analytic model potentials for the atom-surface interactions and induced dipole moments, allowing us to characterize the resulting field fluctuation spectrum for a wide range of atomic or molecular species in terms of a small set of microscopic parameters.   

While more accurate predictions will depend on the atomic species, our model explains correctly the magnitude and the $d^{-4}$ scaling of the observed electric field noise. In contrast to standard models for $1/f$ noise, which assume a distribution of  two level fluctuators, our analysis predicts that for adatoms a transition from a flat to a $1/f$ regime in the noise spectrum should occur only at finite temperatures and at frequencies above a typical phonon transition rate $\omega>\Gamma_0$. From our estimates, we find that this transition rate is in the range, but slightly above the typical ion trapping frequencies. This suggest that either heavier or more loosely bound adatoms are responsible for the noise, or additional mechanism like multi-phonon transitions or dipole-dipole interactions are at play leading to the emergence of even lower fluctuation rates.  

Our model predicts several distinct features for the electric field noise spectrum which appear at characteristics frequency and temperature scales of the adatom surface interactions.   
In future experiments the development of new  trap designs~\cite{NeedleTrap}  could allow for more targeted search for these predictions, e.g. by probing samples where a clean metal surface is contaminated with a single, pre-specified atomic species.  To access the frequency regime of 10 to a few 100 MHz, similar heating experiments could be done with nano-mechanical resonators or carbon nanotubes,  for which accurate optical and electrical readout schemes have been developed (see e.g. Ref.~\cite{NMRReview} and references therein). 

%To explore the frequency regime beyond a few MHz, similar approaches could be used for charged nano-mechanical resonators or superconducting devices. Here   

\section{Acknowledgements}
The authors thank D. Wineland for valuable discussions.  This work was supported by NSF through a grant to ITAMP at the Harvard-Smithsonian Center for Astrophysics.

\end{document}